# Modelling pyro-convection phenomenon during a mega-fire event in Portugal

**Cátia Campos**

Instituto de Ciências da Terra – ICT (Polo de Évora),
Earth Remote Sensing Laboratory (EaRS Lab),
Universidade de Évora.
Rua Romão Ramalho, 59, 7000-671, Évora, Portugal.
E-mail: catia.campos@uevora.pt

**Flavio Tiago Couto**

Instituto de Ciências da Terra – ICT (Polo de Évora),
Instituto de Investigação e Formação Avançada – IIFA,
Earth Remote Sensing Laboratory (EaRS Lab),
Departamento de Física, Escola de Ciências e Tecnologia,
Universidade de Évora.
Rua Romão Ramalho, 59, 7000-671, Évora, Portugal.
E-mail: fcouto@uevora.pt

**Jean-Baptiste Filippi**

Centre National de la Recherche Scientifique (CNRS),
Sciences Pour l'Environnement – Unité Mixte de Recherche 6134,
Università di Corsica, Campus Grossetti, Corte, France.
E-mail : filippi_j@univ-corse.fr

**Roberta Baggio**

Centre National de la Recherche Scientifique (CNRS),
Sciences Pour l'Environnement – Unité Mixte de Recherche 6134,
Università di Corsica, Campus Grossetti, Corte, France.
E-mail: baggio_r@univ-corse.fr

**Rui Salgado**

Instituto de Ciências da Terra – ICT (Polo de Évora),
Instituto de Investigação e Formação Avançada – IIFA,
Earth Remote Sensing Laboratory (EaRS Lab),
Departamento de Física, Escola de Ciências e Tecnologia,
Universidade de Évora.
Rua Romão Ramalho, 59, 7000-671, Évora, Portugal.
E-mail:  rsal@uevora.pt

**Abstract**

The present study contributes to an increased understanding of pyro-convection phenomena by using a fire-atmosphere coupled simulation, and investigates in detail the large-scale meteorological conditions affecting Portugal during the occurrence of multiple mega-fires events on 15 October 2017. Two numerical simulations were performed using the MesoNH atmospheric model. The first simulation, was run for a large single domain (300 × 250 grid points) with a 15 km resolution. In the second one, the MesoNH was coupled to a fire propagation model (ForeFire) to study in detail the Quiaios's fire. To optimize both high resolution in the proximity of the fire region and computational efficiency, the simulation is set up using 3 nested domains (300 × 300 grid points) with horizontal resolution of 2000 m, 400 m, and 80 m respectively. The emission into the atmosphere of the heat and the water vapour fluxes caused by the evolving fire is managed by the ForeFire code. The fire spatio-temporal evolution is based on an assigned map, which follows what reported by public authorities. At the large scale, the simulation shows the evolution of the hurricane Ophelia, pointing out the influence of south/southwest winds on the rapid spread of active fires, as well as the subtropical moisture transport toward mainland Portugal in the early evening, when violent pyro-convective activity was observed in Central Portugal. The coupled simulation allowed to reproduce the formation of a PyroCu cloud inside the smoke plume. The convective updraughts caused by the fire led to the vertical transport of water vapour to higher levels and enhanced the development of a high-based cloud over a dry atmospheric layer within the smoke plume.

# 1. Introduction

Wildfires are the result of complex interactions between biological, meteorological, physical, and social factors that influence the intensity, spread, duration and extent, as well as their potential to cause damages. In recent years, there has been an increase in frequency of very large wildfires events, known referred as mega-fires. These events are characterized with high fire rate of spread, large burnt areas (> 10,000 hectares) and significant negative impacts on the environment and society. For instance, despite being a minority in the global number of fires, the forest area burnt from mega-fires have catastrophic consequences in biodiversity (Bento-Gonçalves, 2021; Duane et al., 2021; Tedim et al., 2020). Countries such as Portugal, Australia, Canada, USA, and Greece experienced these most catastrophic events in recent years (e.g., Couto et al., 2020; Peterson et al., 2021; Ranjbar et al., 2019; Kartsios et al., 2020).

In the last years, strong pyro-convective activity has been increasingly reported during the occurrence of mega-fires, which becomes visible through the formation of clouds from the convection created by the fire itself. The Flammagenitus or cumulonimbus flammagenitus clouds are also known as pyro-cumulus (PyroCu) or pyro-cumulonimbus (PyroCb), respectively (e.g., International cloud atlas, 2022; Fromm et al., 2010; Badlan et al., 2021a). Under such fire environments, several studies identified thermodynamic diagrams showing the temperature (T) and dew point temperature ($T_d$) vertical structure as an inverted V, indicating a dry and stable lower troposphere, followed by a moist layer in the middle levels in which cloud development may occur (Giannaros et al., 2022; Couto et al., 2020; Tory et al., 2018; Peterson et al., 2017; Laureau and Clements, 2016; Rosenfeld et al., 2007; Goens and Andrews, 1998).

The simplest form of these pyro clouds is the PyroCu, which extends above the lifting condensation level (LCL) to the middle troposphere. However, when the smoke column is large enough, it can be reinforced by the potential instability in the middle troposphere. In this case, the convective column penetrates the upper troposphere or even the lower stratosphere, leading to the formation of PyroCb clouds (e.g., Fromm & Servranckx, 2003; Badlan et al., 2021a; Duane et al., 2021).

Such clouds are fire driven, associated with strong sensible heat flux on the ground as well as injection of moisture into the plume that can contribute to the PyroCb development by additional moist air and latent heat (Giannaros et al., 2022; Badlan et al.,

2021a; Thurston et al., 2015). Due to the cloud vertical extent, PyroCb clouds are capable of injecting ashes and other aerosols into the lower stratosphere, where they can be redistributed around the globe (e.g., Laureau and Clements, 2016; Fromm et al., 2021; Sicard et al., 2019). In terms of microphysical structure, such aerosols in the upper troposphere are referred as pyrometeors and can have long range/distance impact at synoptic scales (e.g., McCarthy et al., 2019). At the meso or micro-scale, PyroCb clouds are characterized by being extremely violent and can trigger extreme phenomena such as fire-whirls or the ignition of new fires from electrical discharges (e.g., Lang et al., 2014; Thurston et al., 2015). Moreover, the winds induced by such phenomena can strongly affect the fire propagation, and may result in an acceleration of the front because of these complex coupled interactions.

Computational advances, permit an increase in forecast resolution, allowing the development of coupled fire-atmosphere models, which helped to understand and anticipate fire behaviour and contributes to the protection of civilians and firefighters. Specifically, simulating the coupled effects between fire and atmosphere have been the subject of many efforts in the last few years (Mandel et al. 2014, Coen et al., 2020; Filippi et al., 2018). Through modelling, it is possible to predict the smoke plume transport, and several physical processes associated with extreme fires, such as sudden wind changes (in terms of intensity and direction) that cause difficulties in firefighting.

Strong convection over the fire head is one of the main fire-to-atmosphere effects. Couto et al. (2022a) simulated such a phenomenon using the MesoNH/ForeFire coupled code. They showed convection being enhanced by the fire activity and penetrating the upper troposphere. Besides the PyroCb cloud, which was represented by five hydrometeors species, the coupled simulation showed microbursts originating from the PyroCb cloud. Furthermore, Couto et al. (2020) showed that violent pyro-convective activity in Pedrógão Grande mega fire occurred under a dry thunderstorm's environment. The development of a mesoscale convective system produced downbursts surrounding the wildfire event. Such a phenomenon occurs when precipitation, generated in the middle troposphere, evaporates as it falls to the surface, the air present in the dry layer cools, becoming denser and accelerates towards the surface (e.g., Thurston et al., 2015). At surface, fire spread can be highly unpredictable and difficult to suppress because of these erratic winds created by downbursts.

In addition to local conditions created in the fire environment, large-scale dynamics can also favour the intensification or suppression of fires. For instance, Portugal is affected by several types of large-scale weather systems, such as high/low pressure systems, extratropical cyclones, or hurricanes. Novo et al. (2022) showed how the synoptic patterns together with the thermal low over Portugal produce critical fire conditions. In the same sense, Purificação et al. (2022) presented the important role played by the large-scale environment favouring the rapid fire spread during the Vila de Rei's fire in July 2019. In 2017, the approach of Ophelia hurricane induced strong winds over already active fires in mainland Portugal (Simões et al., 2018). While hurricanes are not common in mid-latitudes, they are becoming more frequent with some of them that can reach mainland Portuguese territory, causing strong winds and precipitation (e.g., Pasch and Roberts, 2019; Stewart, 2018; Franklin, 2006).

This study aims to analyse in detail the impact of a fire on the atmosphere during a mega-fire occurred on 15 October 2017 using a fire-atmosphere coupled simulation to better understand pyro-convective activity and the complex influence of the fire in the atmosphere at the micro-scale.

The manuscript is organized as follows: section 2 presents the case study choice, data and methodology applied in the study. The results are presented in section 3 followed by discussion and conclusions in sections 4 and 5, respectively.

## 2. Case study, data, and methodology

### 2.1 Case study

In 2017, a total of 11 mega-fires were recorded in Portugal with 117 fatalities, of which 48 occurred on 15 October. The first mega-fires occurred in June (Pedrógão Grande and Góis fires), followed by another mega-fire in late July (Sertã fire), and with the last 8 occurring on 15 October. Table 1 shows the 8 mega-fires recorded on 15 October with the respective burned areas. According to an official report of the Portuguese authorities (CTI report, 2018), extreme pyro-convective activity leading to the development of PyroCb clouds occurred in Pataias, Arganil, Lousã and Sertã mega-fires. Among the 8 episodes, the Quiaios mega-fire was selected as a case study.

According to the IPMA report (Simões et al., 2018), the 15 October was characterized by extreme fire weather conditions associated with a low moisture content of fuels, resulting in high values of the meteorological indices. For instance, the Fire Weather Index (FWI) showed exceptionally high values for October, with values above the $50^{th}$ percentile during the first fortnight. It reached the maximum value since 1999 on October 15, 2017 (FWI = 59.2). Regarding the Initial Spread Index (ISI), the index presented value greater than the $90^{th}$ percentile on that day (Simões et al., 2018).

The parish of Quiaios belongs to the municipality of Figueira da Foz (district of Coimbra) with around 3000 residents and it is located near the west coast of Portugal (red circle, Figure 1). The region has a relatively flat orography with altitudes around 50 m (Figure 1b). Concerning land cover, it was composed mainly by forest cover (92.16 %), of which was shrubs (2.31 %) and forest (89.85 %), the latter consisting mainly of maritime pine (64.04 %) and eucalyptus (17.89 %). The agriculture (6.78 %) and urban area (1.06 %) represented a small percentage of land cover (cf. Table 3.2 of CTI report, 2018).

The Quiaios's fire origin was intentional, it started on 15 October at 2:36 pm and being suppressed at 11:10 pm on 16 October. The total burned area reached 19,000 ha, because of several ignitions (Viegas et al., 2019). For instance, the first ignition at 1:34 pm, in Quintã (Vagos), it was followed by the most important ignition in Cova da Serpe, at 2:36 pm (Viegas et al., 2019; CTI report, 2018).

Secondary ignitions with unknown source were also reported and contributed to the total burned area. It is noteworthy to mention a high rate of spread with speed of 4.8 km/h during the first 4 hours, as well as a maximum of 5.4 km/h (CTI report, 2018). Although official reports, focused on fire behaviour, did not mention intense pyro-convective activity, some photographs show evidence of a pyro-cumulus cloud development (Foz ao Minuto, 2017).

## 2.2 Meteorological dataset

Meteorological data covering the period of study were used to characterize the local atmospheric conditions and to validate the model outputs.

### 2.2.1 Weather station data

The most critical fire weather variables, i.e., air temperature (°C), relative humidity (%), wind intensity (m/s) and wind direction (°) were obtained from the 4 weather stations closest to Quiaios and provided by the Portuguese Institute for Sea and Atmosphere, I. P. (IPMA, IP). The weather stations correspond to "Vila Verde" (Figueira da Foz), "Dunas de Mira", "Becanta" (Coimbra) and "Coimbra" (aerodrome) stations (Figure 1a). It is important to mention that "Becanta" and "Dunas de Mira" stations did not present wind data. In addition, the climatic normal obtained from the monthly climatological bulletins for the period [1971-2000] (IPMA, 2017) was analysed for two variables, namely the monthly average calculated from the average daily temperature and the monthly average relative humidity at 0900 UTC, both for the month of October and "Becanta" and "Dunas de Mira" stations.

### 2.2.2 Sounding data

To validate the vertical structure of the atmosphere simulated by the model, the study used the data recorded by the sounding lunched in Lisbon at 1200 UTC on October 15, 2017 (University of Wyoming, 2017).

### 2.2.3 Sea surface temperature and remote sensing data

Monthly sea surface temperature (SST) data was obtained from COBE-SST (Ishii et al., 2005; Japan Meteorological Agency, 2006). This data is produced by the Japan Meteorological Agency (JMA) climate assimilation system with a temporal coverage from January 1981 to July 2022 and a global grid with 1° latitude by 1° longitude resolution (JMA, 2022). In the study, the data was extracted for the October months and for the period between 1987 and 2017. In addition to the SST data, a RADAR image from the Global Precipitation Measurement (GPM) observatory are also analyzed (GPM, 2017).

### 2.3 Numerical Modelling

The study is based on two simulations performed with the MesoNH model (Lac et al., 2018). The MesoNH is a limited area and non-hydrostatic atmospheric model, which allows nesting grids and the activation of different types of parameterizations schemes to represent several physical processes in the Earth's atmosphere. In Portugal, it has been successfully used in several research fields, including wildfires research (e.g., Couto et al., 2020; 2021; 2022a; 2022b).

In the study, the fire propagation model ForeFire (Filippi et al., 2009; 2011) is coupled to the MesoNH model. ForeFire allows the computation of the temporal evolution of the fire front line and the emission of energy and mass fluxes from the fire into the atmosphere. The model considers the terrain slope, atmospheric properties (wind speed, air density and air temperature), spatial characterization (mass load, height, emissivity, moisture content) and combustion characteristics of fuels (e.g., ignition temperature, calorific value and enthalpy of combustion), and assumes that the fire propagates in the forward direction normal to the frontline.

The coupled code allows the exchange of fluxes in both directions: the atmospheric model provides the wind field and other meteorological variables to the fire propagation model, while the fire model provides the heat, water vapour, and potentially aerosols and chemical species fluxes into the atmospheric model. The MesoNH/ForeFire code has been successfully used both in the wildfire research (Filippi et al., 2018) and, more recently, exploring the development of a large industrial fire plume (Baggio et al., 2022).

**2.3.1 Numerical experiments**

Two numerical simulations were performed: a large-scale non-coupled simulation and a very-high resolution fire-atmosphere coupled simulation.

**2.3.1.1 Non-coupled simulation: large scale environment.**

To contextualize the local atmospheric conditions on the large-scale environment, an auxiliary large-scale simulation was performed. It is well-known that the synoptic situation on 15 October 2017 was marked by the presence of the hurricane Ophelia in the East coast of the North Atlantic Ocean (e.g., Moore, 2021; Moore, 2019). Therefore, a large domain was designed to follow and characterize the hurricane in all phases of its development. The domain was configured with 300 × 250 grid points and a lower horizontal resolution of 15 km (Figure 2a). The simulation was initialized from the European Centre for Medium-Range Weather Forecasts (ECMWF) analysis (updated every 6 hours) and performed for the period between 0000 UTC on 4 October and 1200 UTC on October 16, 2017. The advantage of using simulation, compared to the use of existing analyses, for example from the ECMWF Integrated Forecasting System (IFS), National Centers for Environmental Prediction (NCEP) / Global Forecast System (GFS), Centre National de Recherches Météorologiques (CNRM) / Action de Recherche Petite Echelle Grande Echelle (ARPEGE), is that the model is resolving all atmospheric dynamics fields of the atmosphere, while the analysis only store the basic variables. The simulation was also configured to produce results with a higher temporal resolution than the available analyses (outputs every 3 hours).

**2.3.1.2 Coupled simulation: local to mesoscale environment.**

To study pyro-convection activity in Quiaios's fire, the Meso-NH/ForeFire coupled simulation was performed over 3 nested domains. Each domain was configured with 300 grid points in the x and y directions (Figure 2b). The largest domain was designed with 2000 m horizontal resolution, whereas the second and third domains with 400 m and 80 m horizontal resolution, respectively. Vertical grid is distributed over 50 model levels, with the level closest to the surface at 30 meters height.

The model run in two segments: first, the large domain of 2000 m resolution run from 0600 UTC to 1300 UTC on 15 October 2017; the model is then restarted at 1300 UTC on 15 October using the 3 two-ways nested domains with only the higher resolution model being coupled with the fire model. The ECMWF operational analyses were used to initialize the model and to provide boundary conditions, updated every 6 hours. The time step configuration was 10 s for the outermost model, decreased to 2 s and 0.5 s for the finest model.

The coupled MesoNH/ForeFire code was applied in the finest 80 m resolution domain to analyse the fire plume evolution and the pyro-convective activity. As stated earlier, ForeFire allows to simulate the propagation of the fire and its impact on the atmosphere. Thanks to the two-way nested configuration, energy and water vapour transferred from ForeFire to Meso-NH in the innermost model are subsequently transported to the other domains.

In this study, the fire started at 1330 UTC and the fire front line evolution was forced using a spatio-temporal evolution map extrapolated from the burned area report presented by Viegas et al. (2019). In the ForeFire model parameterization, fuel is assumed to be homogeneous and set to the fuel model 11 from Anderson (1982) with an average of burning fuel load of 2.5 kgm$^2$ at 30% humidity (0.75 kg of water, 1.75 kg of combustible). Combustion enthalpy is taken at Dh = 1.5 × 10$^7$ J/kg and evaporation Dhw = 2.3 × 10$^6$ J/kg. This results in a total energy released after drying of 2.4525 × 10$^7$ J/m$^2$, but only 40% of this energy is assumed to be sensible fluxes heating the atmosphere, as combustion can be incomplete and most of the energy is lost in radiation toward the ground and the space and to heat the combustible to pyrolysis. The resulting 0.75 kg/m$^2$ of water vapor and 10 × 10$^6$ J/m$^2$ are released in the atmosphere at the location of the moving fire front for a 1000s burn duration, resulting in instantaneous fluxes of 0.75 g/m$^2$s of water vapour and 10 KW/m$^2$. The fire model (heat fluxes computation) was performed on a 20 meters resolution grid (1200 × 1200). Output files were written every 10 s of simulation (80 m), 100 s for the 400 m, and 600 s for the larger model.

**2.3.2 Parameterization schemes used in the experiments.**

The parametrization schemes used in the simulations are shown on Table 2. For all experiments and domains, the microphysical scheme considered was ICE3 (Pinty and Jabouille, 1998; Caniaux et al., 1994). This one-moment bulk scheme allows the simulation of cloud systems, considering up to 6 classes of hydrometeors: water vapour, cloud droplets, raindrops, ice crystals, snow and graupel. The radiation parameterization was based on the Rapid Radiative Transfer Model (Mlawer et al., 1997). For the large-scale simulation, deep convection was parameterized from the KAFR scheme (Bechtold et al., 2001), whereas shallow convection was parameterized using the EDKF scheme (Pergaud et al., 2009). Regarding the parameterization of turbulence, a 1D turbulence scheme (Cuxart et al., 2000) was activated for the larger domains, as the horizontal fluxes are negligible compared to the vertical fluxes at coarse resolutions. In the high-resolution simulation, the full 3D turbulent scheme was activated.

## 3. Results

Results are divided into two subsections: Section 3.1 presents the large-scale features of the weather system, whereas the fire weather conditions obtained from the observations are presented in Section 3.2 jointly with the results of the coupled simulation.

### 3.1 Large-scale environment

Based on literature review it is estimated that the hurricane Ophelia was active on the East coast of the North Atlantic Ocean between 6 and 17 October 2017. Given that the development of such phenomenon in the North Atlantic Ocean is very rare, we explore the main features using the simulation described in Section 2.3.1.1.

The simulation indicated a low-pressure system in mid-latitudes, centered around 46 °N and 42 °W and with mean sea level pressures around 995 hPa at 1200 UTC on 04 October 2017 (Figure 3a). The frontal zone extended from the low-pressure centre towards lower latitudes. The same figure displays a warm airmass with temperatures above 25 °C covering large part of the Atlantic Ocean. The large-scale pattern remains over the region during the first simulated days. However, from October 7, 2017, a change in the wind circulation can be identified along the frontal zone. For instance, Figure 3b shows a low-pressure core involved by a cyclonic circulation at 0600 UTC on 07 October 2017. It is important to mention that this low-pressure core develops within the subtropical warm air mass that remains predominant in the region, as well as over warmer oceanic waters (Figure 4a). Figure 4a displays the sea surface temperature (SST) obtained from the ECMWF Analysis on 5 October at 0000 UTC with SST values between 24 and 28 °C in the region where the system developed ([30 – 40] °N and [40 – 45] °W). On the other hand, the mean SST for a 30-years period [1987 – 2017] for October shows values between 20 °C to 24 °C (Figure 4b). Therefore, the SST during the passage of the frontal system was higher in relation to the climatic mean for October, with a positive anomaly of 2 °C to 4 °C. Such a result help to understand the formation of Ophelia in mid-latitudes and its tropical characteristics.

The system continued its evolution and intensification in the following days and while it was moving toward Portugal. The hurricane reached category 2 in the late afternoon of October 12 (Figure 3c), and category 3 (major hurricane stage) on October

14 (Figure 3d). In general, the model simulated the system developing still under the influence of a warmer airmass (temperatures around 25 °C) and with mean sea level pressure around 990 hPa.

The satellite image presented in Figure 5a displays the system located southward of the Azores archipelago with heavy precipitation around the eyewall (70 to 80 mm/h) on October 14, 2017. The radar reflectivity model output (Figure 5b) shows reflectivity values above 40 dBz at the same region around the eyewall. In addition to the internal structure, the deep vertical structure of the system was assessed from the brightness temperatures output (Figure 5c). The lowest values around 210 K indicate the presence of deep clouds in the eyewall. Higher values, around 230 K, are visible around the hurricane's eye and in a band extending north-eastward from the centre of the system. At this date, simulated cloud structure is in good qualitative agreement with the observation.

The simulation showed the hurricane approaching to mainland Portugal on 15 October. Figure 6a shows Ophelia westward of Portugal at 1500 UTC, as well as south-easterly winds over Portugal. The model simulated intense gusty winds of above 40 m/s in the hurricane's core and below 20 m/s over Portugal. At the same instant, relative weak winds of around 20 m/s are simulated in the upper troposphere (Figure 6b). Later, at 2100 UTC (Figure 6c), Ophelia was centered north-westward of the Iberian Peninsula inducing southwest winds over the Portuguese coast in the lower troposphere. The winds remained relatively weak in the upper troposphere (Figure 6d).

Besides the wind field that influenced the evolution of the active fires in the evening of 15 October, the large-scale simulation also indicates the transport of subtropical moisture towards mainland Portugal. The precipitable water field shows values around 30 mm over the country at 2100 UTC (Figure 7a) and more than 35 mm at 0000 UTC on 16 October (Figure 7b). After this period, the hurricane Ophelia starts its transition into an extratropical cyclone, continuing its journey towards Ireland.

### 3.2 Local conditions and pyro-convective activity at Quiaios's fire

Observed vertical atmospheric structure over Lisbon at 1200 UTC is first compared to the simulation aiming to evaluate the simulation. Figure 8 plots the vertical profiles of air (Figure 8a), potential, virtual potential, and equivalent potential temperatures (Figure 8b), and the representation on the Skew-T diagram (Figure 8c), all for the 2000 m resolution domain. Figures 8a and 8b show overall agreement for the vertical temperature profiles simulation. Considering Figure 8c, the simulated vertical structure of the atmosphere follows a very similar pattern to the observed one: in the layer [1000 - 700] hPa, the atmospheric profiles are also well represented by the model indicating a dry layer in the lower troposphere. The same figure shows a dew point temperature ($T_d$) overestimation by the model in the upper troposphere and tropopause. From 600 hPa, the simulation shows a much more humid atmosphere ($T_d$ close to T) in relation to the observed data. Even with such overestimation, the simulated boundary layer is acceptable to represent the atmospheric state and the simulation can be used to understand the case study. Figure 8c also indicates a "dry thunderstorm environment" on October 15, 2017.

Figure 9 displays the main fire weather variables obtained from the meteorological stations on 15 October 2017. It should be noted that near surface air temperature in the period [1000 – 1600] UTC is always above 30 °C (Figure 9a). The "Becanta" (orange line), and "Dunas de Mira" (purple line) stations recorded the highest temperature at 1500 UTC (35.5 °C and 35.4 °C, respectively). The air temperature decreases by about 10 °C from this time until at 2300 UTC. Regarding relative humidity (Figure 9b), the "Dunas de Mira" meteorological station (purple line) recorded values above 80 % during the night, with "Vila Verde" station (blue line) registering the highest relative humidity value throughout the day, probably because the stations are located near the coast, under more maritime influence. It is notable that during the afternoon, the relative humidity in the "Becanta" and "Coimbra" stations are below 20 %, with a minimum at 1600 UTC of 12 % and 16 %, respectively.

According to the climatological report (IPMA, 2017), the climatic normal of "Dunas de Mira" station has a mean air temperature of 15.4 °C, and mean relative humidity of 87 %. Regarding the "Becanta" station, the normal for the average temperature is 16.6 °C and 82 % for the mean relative humidity. The temperature (relative humidity) recorded on 15 October were much higher (lower) than normal weather conditions.

In relation to wind behaviour, Figure 9c represents the hourly variation of the mean wind intensity (m/s) and Figure 9d the hourly variation of the mean wind direction (°) observed at 10 m. The wind speed increases from the late morning in both meteorological stations, and maximum intensity of 8 m/s and 8.9 m/s are recorded at 1400

UTC in "Coimbra" and "Vila Verde" stations, respectively. In addition to the wind speed intensification, there is a change of wind direction throughout the day: in the morning, the wind is mainly from the Southeast, whereas it becomes Southwest during the mid-afternoon.

Table 3 presents the Mean Error (ME) and Root Mean Squared Error (RMSE) calculated in each station for the 400 m resolution domain and during the afternoon. It is noteworthy that there was not any station in the area covered by the innermost domain. In the case of 2m air temperature, the mean error indicates that the model tended to underestimate the temperature (negative values) with a maximum of RMSE around 4 °C ("Dunas de Mira" station). In general, the relative humidity at 2 meters was overestimated by the model with ME and RMSE around 12.5 %. The mean wind magnitude at 10 meters was well captured by the simulation with errors close to zero and a maximum RMSE of 3.4 m/s. The small number of stations and time steps are insufficient to do a deep validation using statistical methods. However, the simple verification presented indicates that the model is able to simulate quantitatively the values observed, even with some under or overestimation.

Considering the simulation with 80 m resolution, the near surface wind field is displayed in Figure 10. The figure presents the wind vectors and wind gusts at 10 m at three timeframes along the afternoon. It is noteworthy that the horizontal wind presents a change of its direction, from southeast to southwest, agreeing with the meteorological observations and large-scale simulation.

In the model, the ignition of the fire was imposed at 1330 UTC (Figure 10a; red circle). At that time, the predominant wind direction was from the southeast and the 10 m wind gusts were stronger northward of Quiaios (> 20 m/s; red star). The simulated wind pattern remains similar 1 hour later (Figure 10b), however it is already possible to observe the fire front and the disturbance in the wind field due to the fire, with wind gust values above 20 m/s along the fire front, and weaker northward of the domain. From 1700 UTC (Figure 10c), the wind changed its direction becoming a more south-westerly flow. The fire front line moves north-eastward, following the direction of the prevailing wind, and wind gusts above > 25 m/s are simulated over the fire front.

The hourly simulated vertical wind intensity profile (m/s) is displayed in Figure 11a. The vertical profile extending up to 9 km altitude is taken in a point near to the fire ignition (purple square in Figure 1b), and between at 1400 UTC and 1900 UTC. In general, the wind graphs at different time steps follow the same configuration: the wind is weaker near the surface, increasing with altitude, reaching a maximum of around 18 m/s between levels [1000 - 2000] m, above which it weakens again slowly. Figure 11b shows the water vapour mixing ratio at 1400 UTC with the highest values found in the first 1000 m, above 8 g/kg near the surface. This figure suggests that near surface flow contributed to the advection of some moisture content to the fire environment. On the other hand, even with a deep dry atmospheric layer in the lower troposphere, Figure 11c indicates the presence of high precipitable water content, with values around 24 mm surrounding the fire region. This fact, together with the Skew-T diagram, shows that water was present in both mid and high levels.

The model outputs were analysed to characterize and study the development of the smoke plume. Considering the innermost domain (80 m resolution), the model simulates an intense convective column from the fire. Figure 12 displays the 3D smoke plume represented by the relative smoke tracer concentration variable (SVT). The updraughts with velocities of surpassing 25 m/s near 1500 m at 1415 UTC. The effects of such a vertical motion over the fire front is found 15 minutes later, at 1430 UTC, in the 400 m resolution domain.

Figure 13 shows a vertical cross section intersecting the smoke plume at 1430 UTC on October 15, 2017. Figure 13a shows the relative smoke tracer concentration (SVT – unit scalar variable, emitted where and when the fire is present) with the fire front being identified by the highest values near surface (> 0.1). The simulated smoke extends up to the middle atmospheric levels. Figures 13b to 13e highlight the impact of heat and water vapour emission into the atmosphere.

Turbulent kinetic energy field, in Figure 13b, presents the intensity of subscale updrafts and downdrafts due to the existence of eddies. The turbulence intensity is higher near the surface (above 5 $m^2/s^2$), which facilitates the vertical mixing of components within the fire environment, intensifying gusty winds that are affecting fire. It affects the vertical transport of heat, allowing it to reach higher levels of the atmosphere, as well as the transport of smoke and water vapour. It is also important to note the turbulent region near the smoke plume top, with values around 2 $m^2/s^2$.

Figure 13c illustrates the vertical motions within the plume, with blues representing downward motions, and reds, upward motions. The vertical velocity is approximately zero outside the plume, whereas inside it, the vertical motions (up/down) are more intense, with upward motions, with vertical velocity of 2 and 3 m/s, and reaching altitudes above 6 km. The updraughts are weaker than those presented in Figure 12, however they extend over a larger region and reveal the impact of fire on the atmospheric local circulation. The water vapour mixing ratio field (Figure 13d) is also impacted by the presence of the fire. The field is strongly disturbed over the fire here, and moisture is found in the higher levels within the plume region. Therefore, the turbulent fluxes and the organized updraughts allowed the vertical transport of water vapour to higher levels. At the smoke plume top there is a significant water vapour content, which partially condenses, giving rise to other hydrometeors species. Figure 13e displays the graupel mixing ratio field and, even with the very low values, the presence of hydrometeors within the convective column confirms that the coupling between the fire propagation model and the atmospheric model can represent the pyro-convective activity leading to the development of pyro clouds. The presence of hydrometeors was found only in the 400 m resolution domain, and then it occurs outside the innermost domain.

Figure 14a shows a picture of the PyroCu cloud observed during the Quiaios's fire and Figure 14b the photographer location. The simulation well represented the northward oriented smoke plume (Figure 14c). Figure 14c also shows the presence of liquid water and frozen water inside the pyro-cloud. Although the concentrations of hydrometeors are low, of the order of magnitude shown in Figure 13e, they are sufficient to indicate the occurrence of microphysical processes inside the fire generated-plume.

It appears that the wind direction forced the smoke plume growth to northward and shifted the transport of moisture into the plume, in addition to the heat and water vapour fluxes injected by the ForeFire model. Such an environment led to the formation of different hydrometeor species, namely cloud droplets and graupel particles.

## 4. Discussion

This study aimed to study pyro-convective activity in Quiaios's fire, as well as identify the atmospheric elements that favoured the evolution of active fires and strong pyro-convective activity in the evening of October 15, 2017.

The hurricane Ophelia was a remarkable meteorological system during the period with several studies highlighting the Ophelia's influence on the transport of mineral dust from the Sahara Desert and smoke particles from the Portuguese fires toward higher latitudes (e.g., Akritidis et al., 2020; Moore, 2019).

In the study, the mean sea level pressure field analysis obtained from a large-scale simulation allowed the identification of an extratropical cyclone with the associated cold frontal system extending up to lower latitudes in early October. The model showed that Ophelia developed from this cold front, consistent with what reported by Stewart (2018). Stull (2017) identified cold fronts as one of the synoptic triggers for the development of tropical lows. In the case of hurricane Ophelia, the cold front did not reach the equatorial region, and formed at a very northern latitude – the tropical low documented in Stewart (2018) was centered at 31.8 °N, whereas in the simulation was approximately 32 °N. The typical latitudes for tropical cyclogenesis are between [5 – 20] °N, where the temperature is favourable for hurricane development (e.g., NOAA, 2021; Wang et al., 2017; Stull, 2017; Ciasto et al., 2016). Outside these latitudes, the SST is usually lower. The present study showed that the SST in October 2017 was higher in relation to the average for the period between 1987 and 2017, with a positive anomaly of about 4 °C in the region where Ophelia developed.

The hurricane Ophelia approached to Portugal in the evening of 15 October, inducing south/southwest winds over land. The most interesting result obtained from the large-scale experiment was the transport of subtropical moisture towards mainland Portugal from the early evening. The meridional water vapour transport in the North Atlantic Ocean is well known by inducing heavy precipitation events in Madeira Island due to orographic effects (Couto et al., 2012, 2015). Here, the pyro-convection was a crucial mechanism for the vertical transport of water vapour and the development of PyroCb clouds. The strong fire-activity occurred in the end of 15 October was associated with intense PyroCb activity, however, the detailed study about these PyroCb episodes fall out of the main goal of the study. Finally, Ophelia started its extratropical transition in the end of 15 October, becoming an extratropical cyclone on October 16, 2017 (Rantanen et al., 2020; Moore, 2019; Stewart, 2018).

For the Quiaios's fire, the MesoNH was coupled to the ForeFire model and from a qualitative evaluation, the model allowed to represent the atmospheric vertical structure and pyro-convection effects on the atmosphere over the plume. At local scale, the atmosphere at surface showed temperatures above 30 °C and relative humidity below 30 % in the early afternoon, conditions that were unexpected for this time of the year. The wind flow was predominant from the southeast during the period. Regarding the vertical structure of the atmosphere, a dry and warmer lower troposphere was followed by a moist layer in the middle levels. The atmospheric conditions documented in this study are like those identified in previous studies about PyroCu/PyroCb development, i.e., an atmospheric profile with a typical inverted V configuration, or a "dry thunderstorm environment" (Giannaros et al., 2022; Tory et al., 2018; Peterson et al., 2017; Laureau and Clements, 2016; Rosenfeld et al., 2007; Goens and Andrews, 1998). Under this vertical structure, it is unlikely that air parcels reach the LCL due to its own buoyancy, but the fires can provide the energy needed for the air to rise. The high LCL indicates that clouds may develop, but with a high base. The coupled code simulated turbulent fluxes and updraughts created by the fire, which favoured the vertical transport of moisture from the surface to above the LCL. The simulation showed a water vapour mixing ratio vertical profile with values around 8 g/kg near the surface. These values are lower than the values observed during extreme precipitation events, e.g., vapour mixing ratio above 15 g/kg (cf. Figure 6a in Couto et al., 2016), however, it probably influenced the injection of moisture into the plume, in addition to the water vapor evaporated by the fuel during combustion.

According to the reports about the mega-fires of October 2017 (CTI report, 2018; Simões et al., 2018), there was no evidence of strong pyro-convective activity in the Quiaios's fire. However, the present study showed the development of a PyroCu cloud during the event. In the simulation, the fire was imposed from 1330 UTC (2:30 pm local time), and the coupled simulation allowed to observe the vertical transport of water vapour within the fire plume and the formation of cloud droplets and graupel particles, i.e., a pyro-convective cloud. The water production is a critical factor in simulating pyro clouds (Cunningham and Reeder, 2009), which are highly intensified by the updraughts (Chang et al., 2015). Badlan et al. (2021a, 2021b) examined how the size and shape of the fire affect pyro-convection. For instance, the larger diameter fires and fire power tend to produce intense convection extending throughout the troposphere or even penetrating the lower stratosphere, also characterizing the development of a PyroCb cloud. In the present study, the results are more consistent with the smaller diameter fires showed by the authors, when the fire produces small clouds which do not develop sufficiently to form organized pyro-convection and extend from the LCL into the middle troposphere.

The simulation well represented the northward oriented smoke plume, as well as the change in the wind direction at surface, from southeast in the early afternoon to southwest from 1700 UTC.

Figure 15 summarizes the main results based on a schematic model for the local and for the large-scale environments analysed in this study, Figure 15a and Figure 15b, respectively. Figure 15a shows the emission of heat and water vapour fluxes from the fire and the ascending air into the smoke plume. The vertical wind profile analysis showed weak winds near the surface and a maximum intensity of 18 m/s at around 1 km altitude. Above this level, the wind intensity decreases again, conditioning the growth of the plume in altitude. The southerly flow produced a northward oriented smoke plume and the event developed under an atmosphere characterized as "dry thunderstorm environment". Such a vertical structure favours the development of high-based clouds (above 3 km altitude) with the fire providing enough energy so that air parcels can reach the LCL and initiate the condensation processes and cloud droplets development. In addition, the lower temperatures above the isotherm of 0 °C favour the freezing process, originating graupel particles. In the large-scale context (Figure 15b), the trajectory of the hurricane Ophelia and the strong winds associated to the system were crucial for the intensification of the active fires in the late afternoon on October 15, 2017.

Hurricane winds by themselves favoured the rapid spread of the fires and led to several mega-fires. The south-westerly flow affecting the Western Iberian Peninsula transported subtropical moist air toward Portugal, intensifying the pyro-convective activity, developing PyroCb clouds.

## 5. Conclusions

The present study aimed to analyse the pyro-convection phenomenon and the atmospheric conditions during a mega-fire outbreak based on numerical modelling.

The large-scale simulation showed the hurricane Ophelia developing from a cold front and over oceanic waters warmer than the expected for the season. The model reproduced the evolution and trajectory of the hurricane and showed how it induced changes in wind direction and created favourable conditions for the rapid spread of the active fires in mainland Portugal. Such an environment brought strong winds and subtropical moisture to the continent, which possibly contributed to the strong pyro-convective activity observed in the other mega-fires. It is important to note that the MesoNH model well simulated the Ophelia's development, even with lower spatial resolution.

Concerning the Quiaios's fire, the episode occurred under favourable fire weather conditions identified from meteorological observations at surface. The study showed the benefits of coupling an atmospheric model to a fire propagation model. The fire front line evolution was deduced from the official reports (forced fire), and the effects of the fire in the atmosphere were analysed considering the emission of heat and water vapour fluxes by the ForeFire model. The simulation showed the wind changing from southeast to southwest throughout the October 15, 2017. Under a dry thunderstorm environment, the simulation at very-high temporal and spatial resolution successfully represented the pyro-convection phenomenon, which was characterized by a northward oriented smoke plume and mainly by the development of a Pyro-Cumulus cloud.

Finally, despite the advances in the wildfire numerical modelling presented in this study, some topics remain open to future studies. For example, simulations aiming to represent the role played by the atmosphere in the evolution of the fire front line is essential to a better representation of the pyro-convective activity associated with mega-fires, and then simulations with a fire propagation model coupled to the atmosphere in a two-way mode is highly recommended. In the large-scale context, and due the lack of studies about hurricanes developing from a cold front in middle-latitudes, the use of high-resolution simulations is also suggested to obtain a more detailed analysis about the Ophelia cyclogenesis, which fall out of the main goal of the present study. Since the impact of these systems over ocean and land, such a study could contribute to the identification of factors leading to their development and a better forecast. Overall, the study of similar events is recommended to increase the knowledge about the important role that weather systems may play on mega-fires development.

**CRediT authorship contribution statement**

**Cátia Campos:** Conceptualization, Data-curation, Methodology, Software, Validation Visualization, Formal-analysis, Investigation, Writing-original-draft. **Flavio T. Couto:** Conceptualization, Methodology, Software, Visualization, Investigation, Writing - Review & Editing, Supervision, Project administration, Funding acquisition. **Jean-Baptiste Filippi:** Software, Resources, Data-curation, Writing - Review & Editing. **Roberta Baggio:** Software, Data-curation, Writing - Review & Editing. **Rui Salgado:** Resources, Writing - Review & Editing, Project administration, Funding acquisition.

**Declaration of Competing Interest**

The authors declare that they have no known competing financial interests or personal relationships that could have appeared to influence the work reported in this paper.

**Acknowledgments**

This research was funded by the national funds through FCT - Foundation for Science and Technology, I.P. under the PyroC.pt project (Refs. PCIF/MPG/0175/2019), ICT project (Refs. UIDB/04683/2020 and UIDP/04683/2020) and also by European Union through the European Regional Development Fund in the framework of the Interreg V A Spain - Portugal program (POCTEP) through the CILIFO project (Ref.: 0753-CILIFO-5-E), FIREPOCTEP project (0756-FIREPOCTEP-6-E), RH.VITA project (ALT20-05-3559-FSE-000074) and H2020 FIRE-RES (GA: 101037419). The authors are grateful to the Portuguese Institute for Sea and Atmosphere (IPMA) for providing meteorological data, to the European Centre for Medium-Range Weather Forecasts (ECMWF) for the provided meteorological analysis and the supercomputing centre of the University of Corsica for their support on the computation of the coupled simulation.

**Figures Captions**

**Figure 1:** a) meteorological stations (stars) and the location of Quiaios (red circle) and b) location of Quiaios (red circle), the fire ignition point (red star) and the point relative to the vertical wind profile (purple square). Orography (m) represented in shading, obtained from the SRTM database.

**Figure 2:** Configuration of the horizontal domains with orography in shaded (m) obtained from the SRTM database: a) non-coupled simulation, with a resolution of 15 km (350 × 250 grid points) and b) coupled simulation, with domains of 300 × 300 grid point and horizontal resolution of: 2000 m (red square), 400 m (dashed blue square) and 80 m (blue square); the location of Quiaios is represented with a red circle in the innermost domain.

**Figure 3:** Meteorological environment obtain from 15 km domain: a) air temperature at surface (shading, ºC), wind at 10 m (vectors, m/s) and mean sea level pressure (isolines, hPa) for a) October 04$^{th}$, 2017, at 1200 UTC, b) October 07$^{th}$, 2017, at 0600 UTC, c) October 12$^{th}$, 2017, at 1800 UTC, and d) October 14$^{th}$, 2017, at 1200 UTC.

**Figure 4:** Sea surface temperature (shading, °C): a) obtained from ECMWF analysis for October 5$^{th}$, 2017, at 0000 UTC and b) obtained from of COBE-SST2, relative to the monthly average for all October months of the period [1987 – 2017].

**Figure 5:** a) The GPM Microwave Imager (GMI) and Dual-Frequency Precipitation Radar (DPR) instruments recorded data that showed the locations of heavy precipitation associated with hurricane Ophelia for October 14, 2017, at 1656 UTC (Available online at: https://www.nasa.gov/feature/goddard/2017/ophelia-atlantic-ocean), b) zoom from the 15 km resolution domain of the RADAR reflectivity (shading, dBz) simulated by MesoNH and mean sea level pressure (isoline, hPa) on October 14$^{th}$, 2017, at 1500 UTC, and c) cloud brightness temperature (shading, K) simulated by MesoNH and mean sea level pressure (isoline, hPa) on October 14$^{th}$, 2017, at 1500 UTC, for the 15 km domain resolution.

**Figure 6:** a) wind gusts (shading, m/s) and wind at 10 m (vectors, m/s) at 1500 UTC, b) wind intensity and wind vectors at 250 hPa at 1500 UTC, c) wind intensity (shaded, m/s) at 850 hPa and wind at 10 m (vectors, m/s) at 2100 UTC, and d) wind intensity and wind vectors at 250 hPa at 1500 UTC.

**Figure 7:** Precipitable water content (shading, mm) a) at 2100 UTC, and b) at 0000 UTC.

**Figure 8:** Comparison between simulation (MNH) and observations (OBS) for Lisbon at 1200 UTC at coordinate points [38.76 º N 9.13 º W]: a) air temperature (º C), b) potential, equivalent potential, and virtual potential temperatures (K) and c) Skew-T diagram with dew point temperature $T_d$ (DP) and air temperature (T) variation.

**Figure 9:** Fire weather variables on 15 October 2017 for the meteorological stations of "Vila Verde", "Becanta", "Coimbra" and "Dunas de Mira": a) mean air temperature (º C) at 1.5 m, b) mean relative humidity (%) at 1.5 m, c) mean wind intensity (m/s) and d) mean wind direction (º ), with the dashed-line boxes representing the prevailing wind direction for each station, whereas the right side of the Y-axis represents their respective quadrants.

**Figure 10:** Wind gusts (shading, m/s), wind at 10 m (vectors, m/s) for October 15, 2017, at: a) 1330 UTC, b) 1430 UTC and c) 1700 UTC in the 80 m resolution domain. The fire front is represented by the red contour, the location of Quiaios by red circle and the fire ignition point by the red star.

**Figure 11:** a) hourly vertical wind profiles (m/s), b) vertical profile of the mixing ratio vapor (g/kg) at 1400 UTC. Both vertical profiles are obtained for the coordinate point [40.209 º N 8.829 º W]. c) precipitable water content (shading, mm), with the location of Quiaios (red circle) on October 15, 2017, at 1430 UTC, in the 400 m resolution domain.

**Figure 12:** 3D smoke plume (shading), represented by the relative smoke tracer concentration and wind intensity (m/s) on October 15, 2017, at 1415 UTC, in the 80 m resolution domain.

**Figure 13:** Vertical cross section intersecting the plume at 1430 UTC on 15$^{th}$ October 2017: a) relative smoke tracer concentration field, represented by the variable SVT (shading), b) turbulent kinetic energy (shading, m$^2$/s$^2$), c) vertical velocity (shading, m/s), d) water vapor mixing ratio (shading, kg/kg) and e) graupel mixing ratio, (shading, kg/kg). East view of the smoke plume from the 400 m resolution domain.

**Figure 14:** a) photograph of the Quiaios fire on October 15, 2017, (available at the link: http://www.fozaominuto.com/2017/10/incendios-no-concelho-da-figueira-da_15.html), b) location of the photographer and c) simulation obtained by coupling the MesoNH and ForeFire models. Smoke variable represented by the variable SVT (shading), wind at surface (vectors) and graupel (dark blue) and raindrops (light blue), on October 15, 2017, at 1430 UTC, for the 400m resolution domain.

**Figure 15:** Conceptual scheme: a) local environment of the Quiaios mega-fire and b) large-scale environment and mega-fires.

**Tables Captions**

**Table 1:** Mega-fires verified over the weekend [14-16] of October 2017. Source: CTI (2018).

**Table 2:** Summary table of the parameterizations used in this study.

**Table 3:** Mean Error (ME) and Root Mean Squared Error (RMSE) for the 2m air temperature (T2M), 2m relative humidity (RH2M), and mean wind magnitude at 10 meters (WIND). The scores are calculated for the nearest point of each station in the 400 m resolution domain.

## Declaration of Competing Interest

The authors declare that they have no known competing financial interests or personal relationships that could have appeared to influence the work reported in this paper.

CRediT authorship contribution statement

**Cátia Campos:** Conceptualization, Data-curation, Methodology, Software, Validation Visualization, Formal-analysis, Investigation, Writing-original-draft. **Flavio T. Couto:** Conceptualization, Methodology, Software, Visualization, Investigation, Writing - Review & Editing, Supervision, Project administration, Funding acquisition. **Jean-Baptiste Filippi:** Software, Resources, Data-curation, Writing - Review & Editing. **Roberta Baggio:** Software, Data-curation, Writing - Review & Editing. **Rui Salgado:** Resources, Writing - Review & Editing, Project administration, Funding acquisition.

**Declaration of Competing Interest**

The authors declare that they have no known competing financial interests or personal relationships that could have appeared to influence the work reported in this paper.

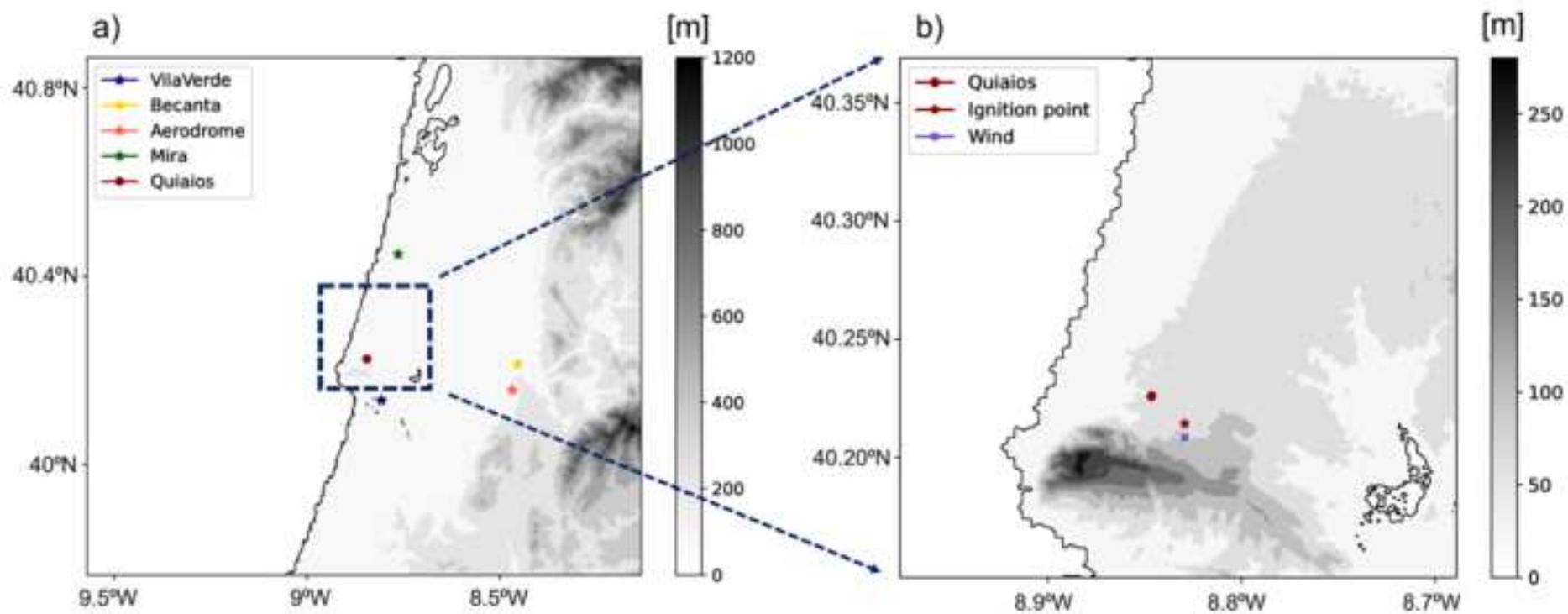

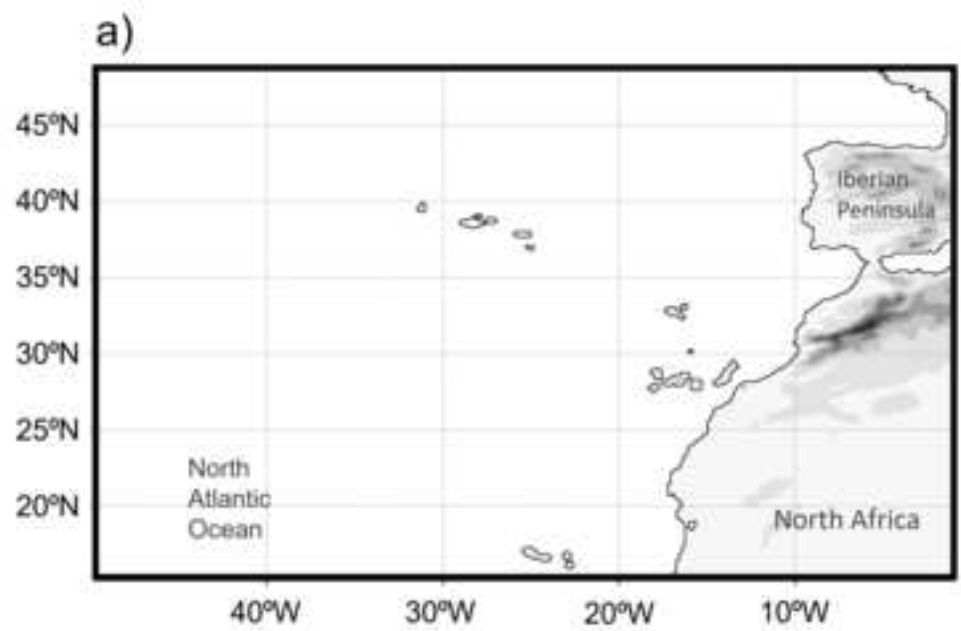
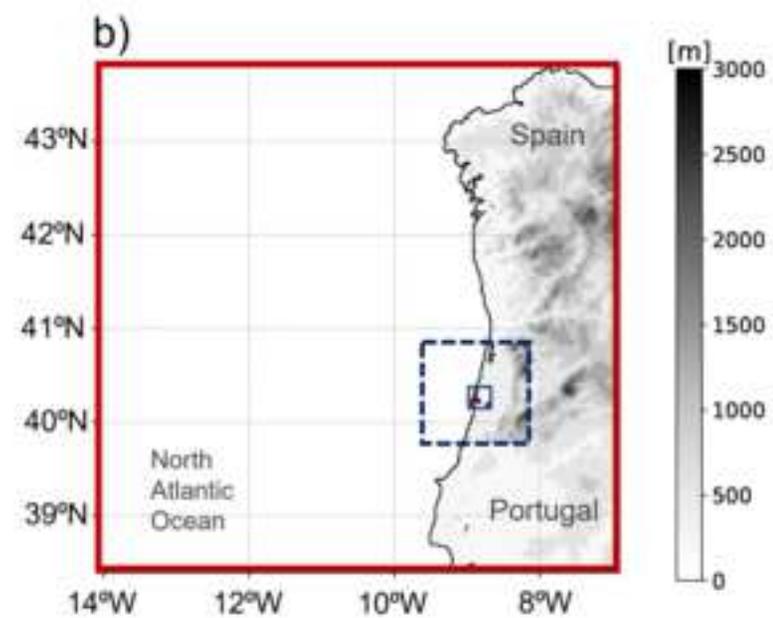

Figure_03

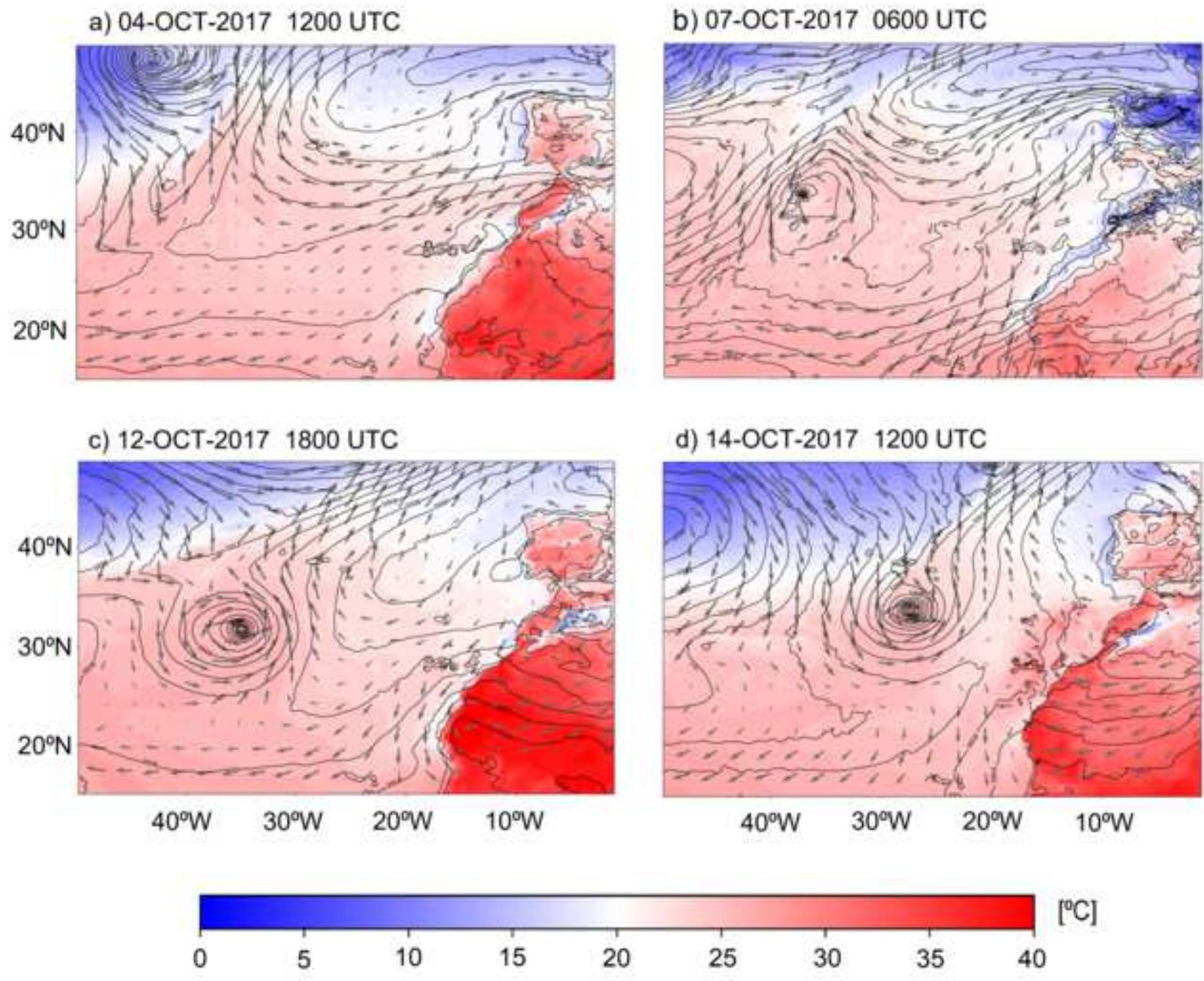

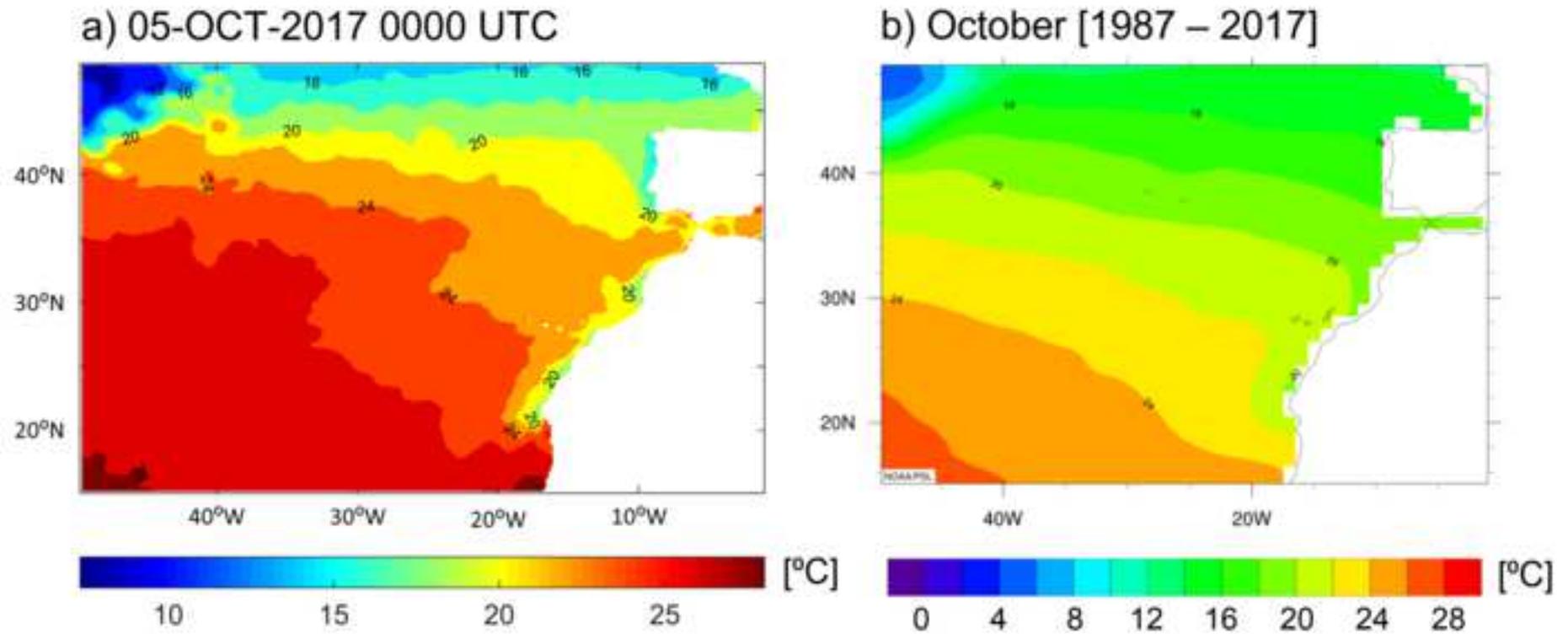

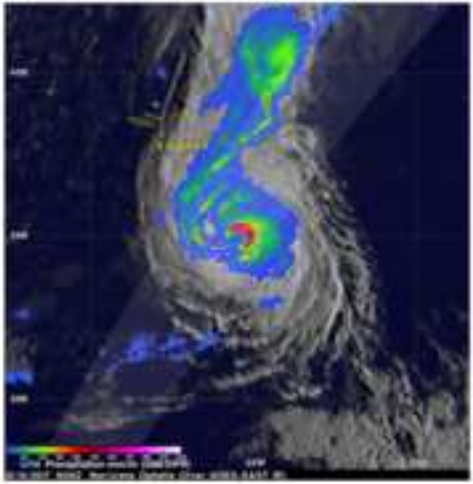
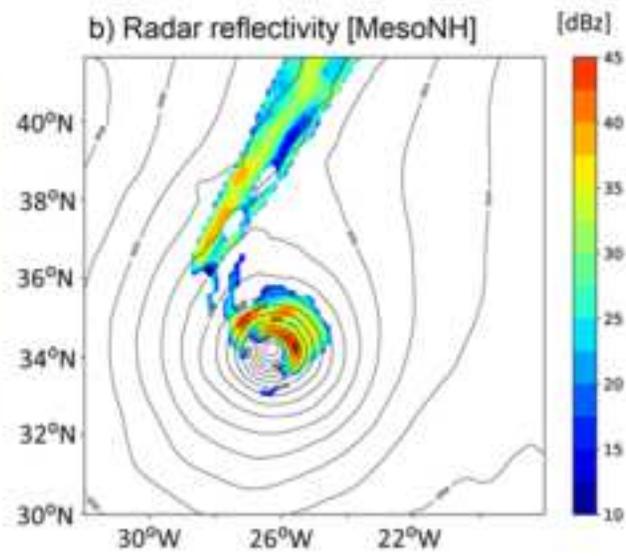
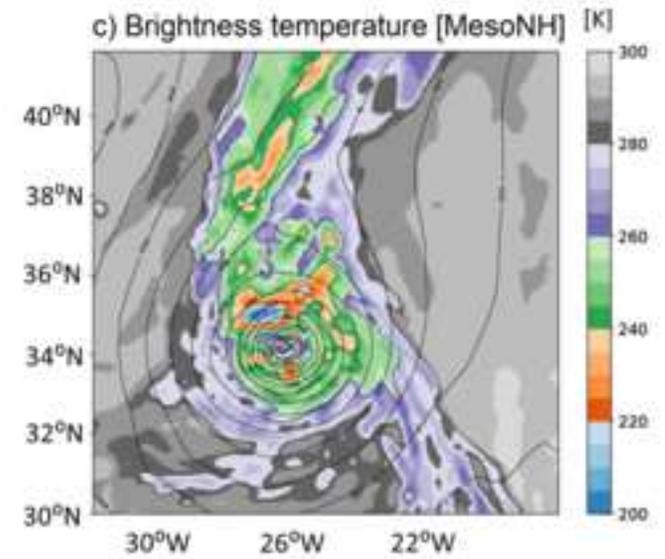

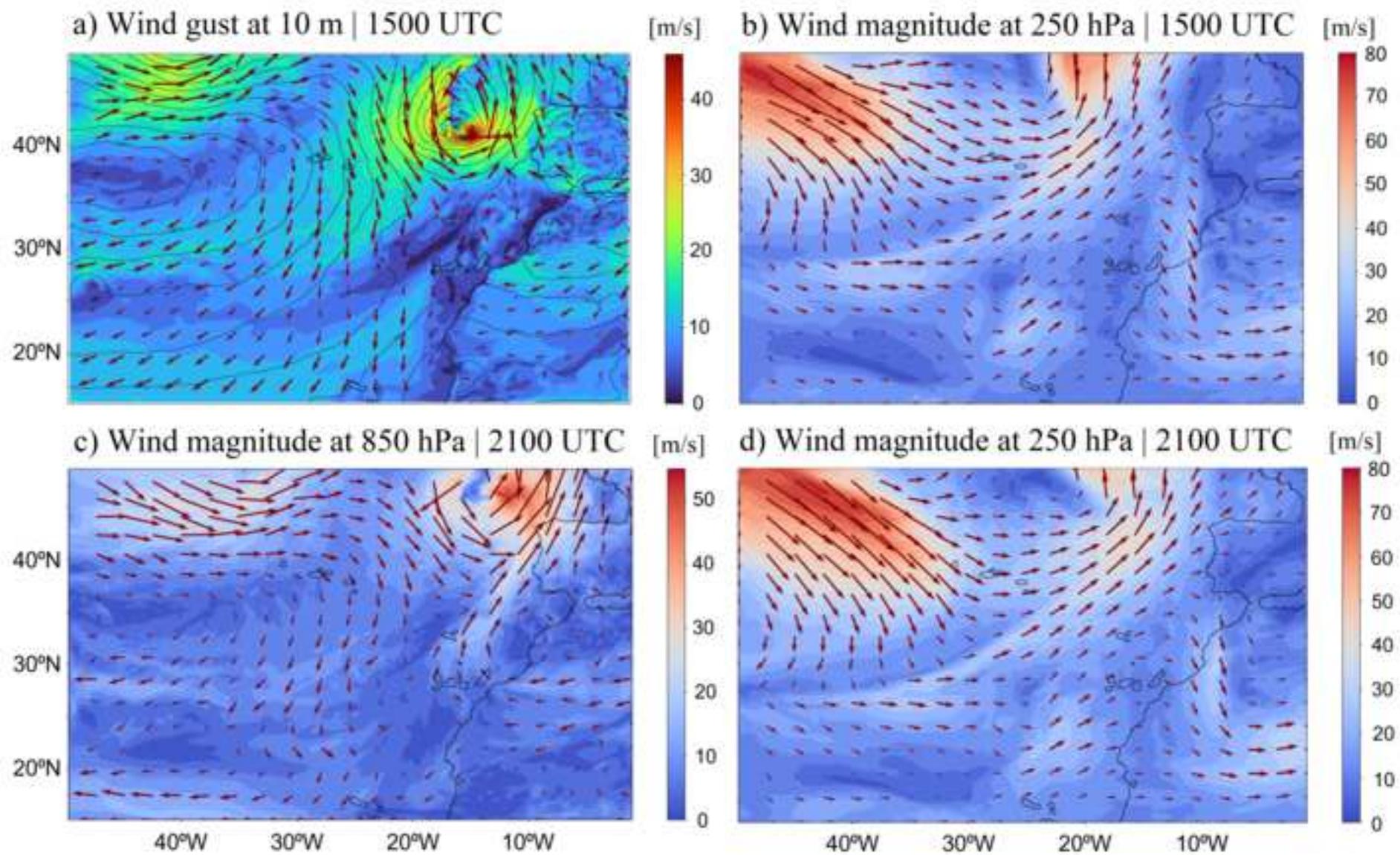

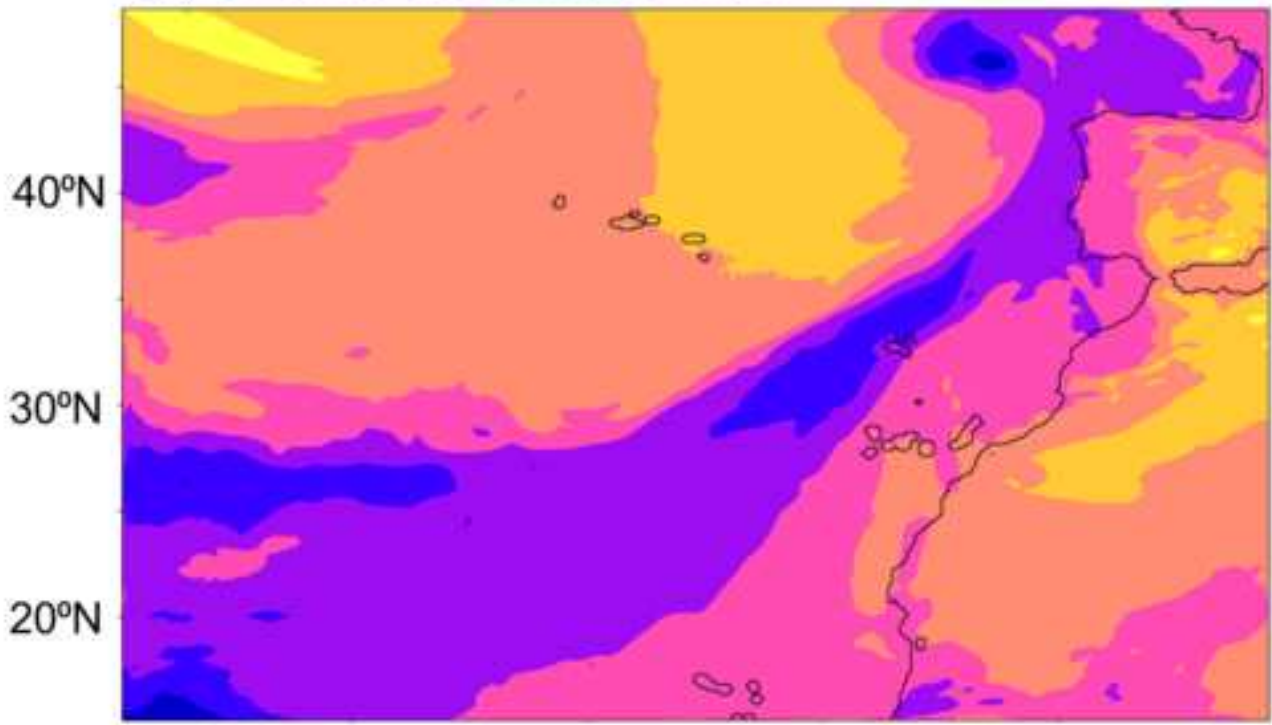
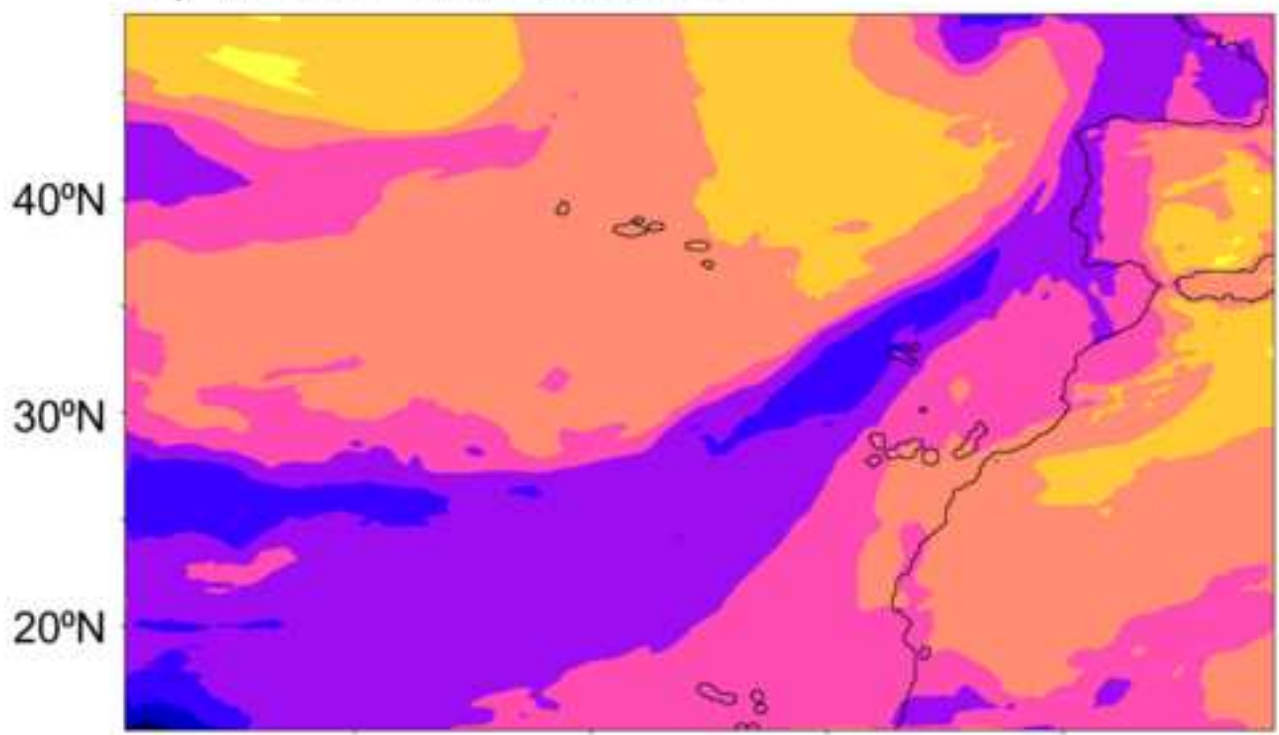
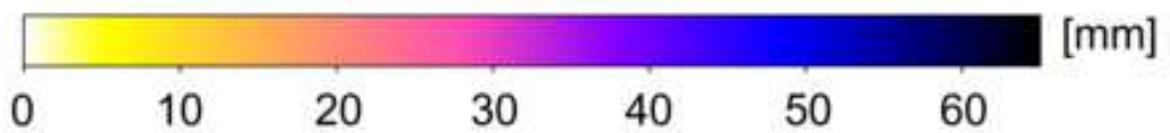

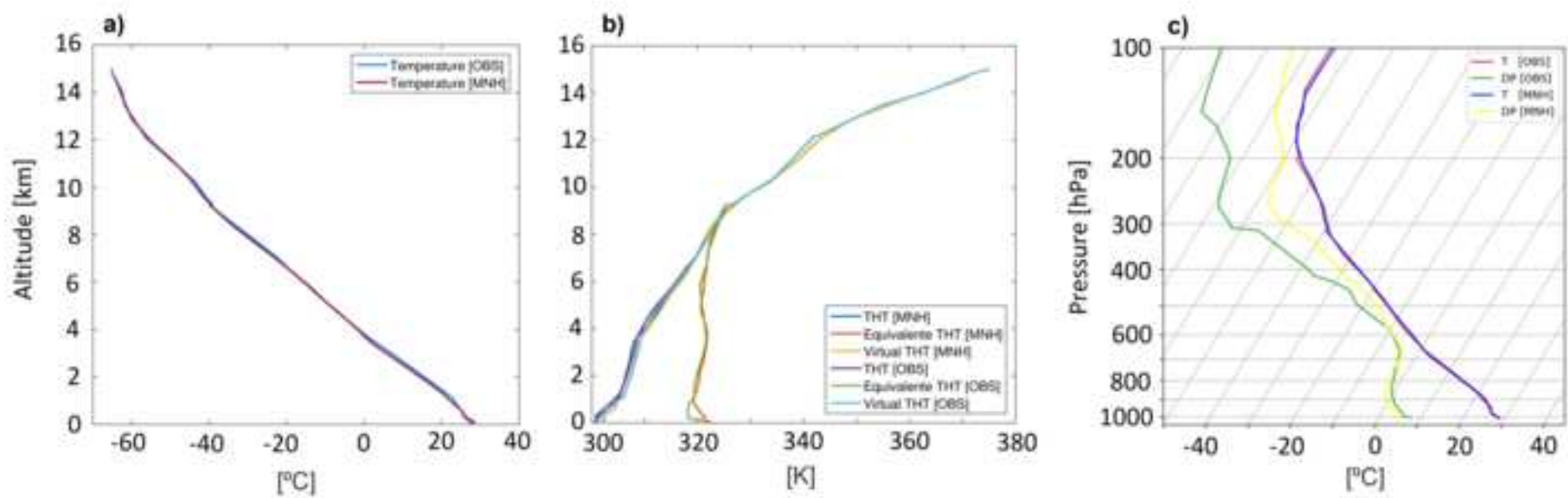

Figure_09

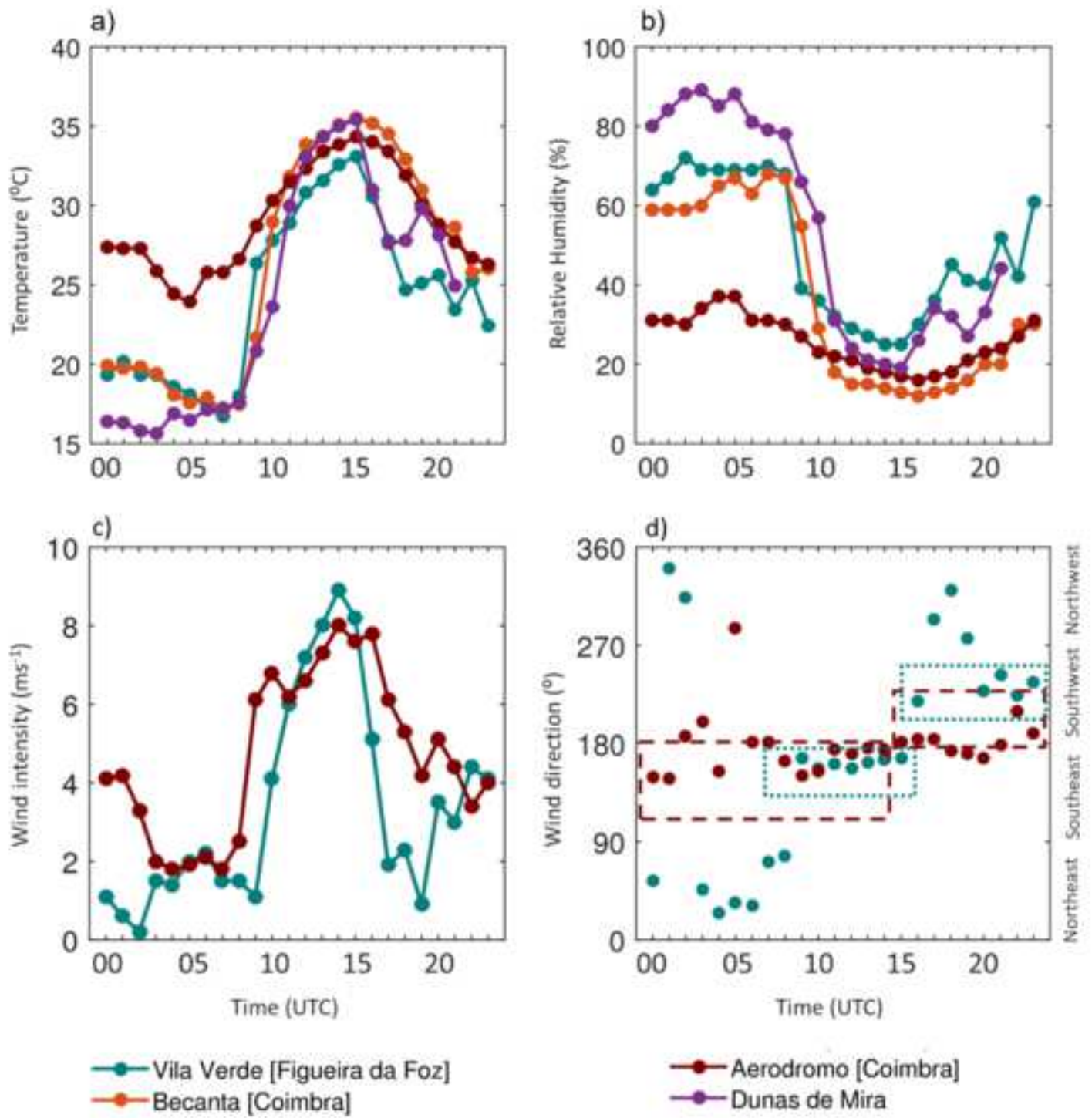

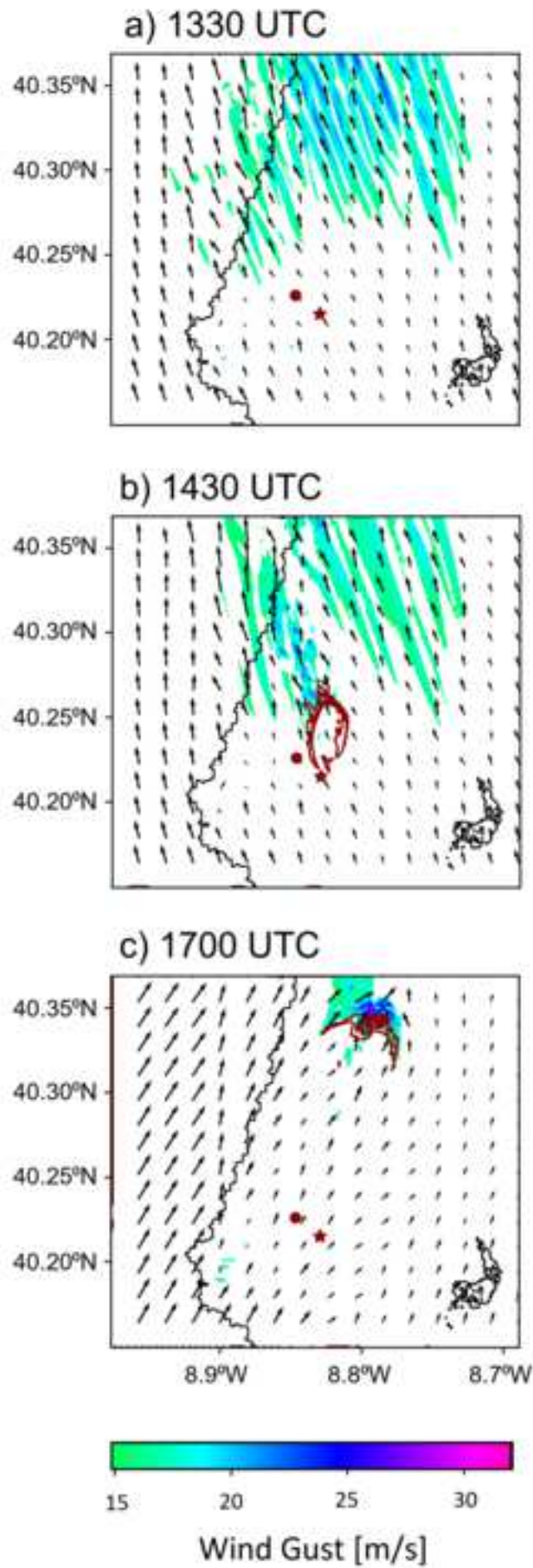

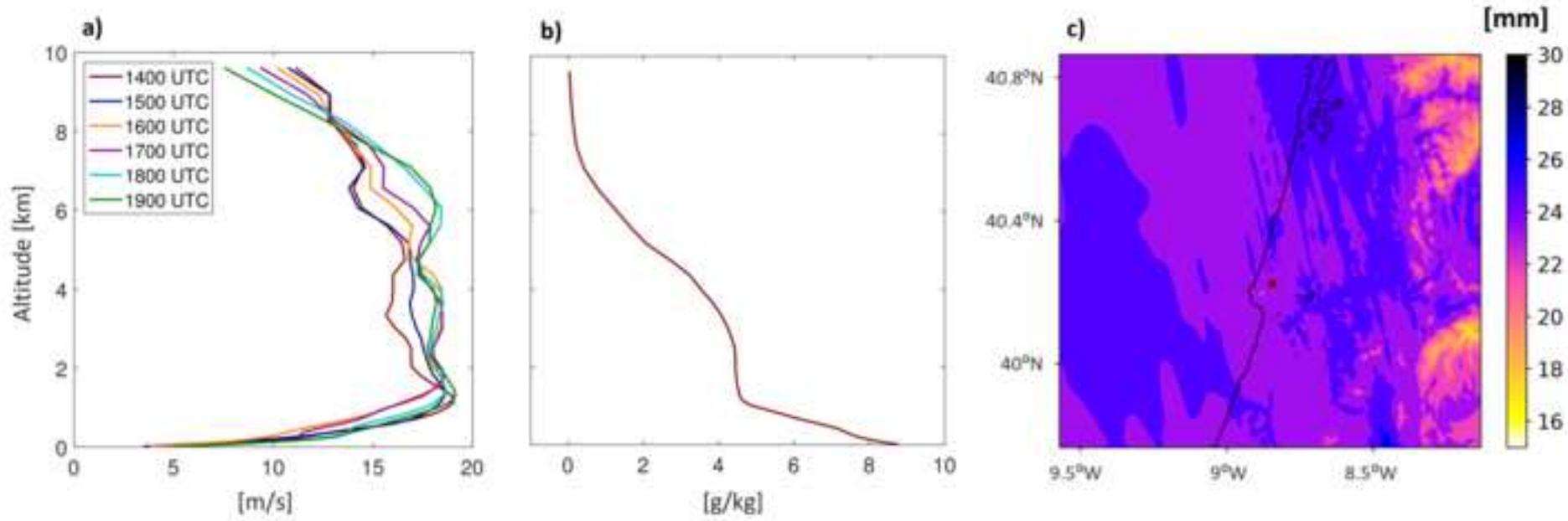

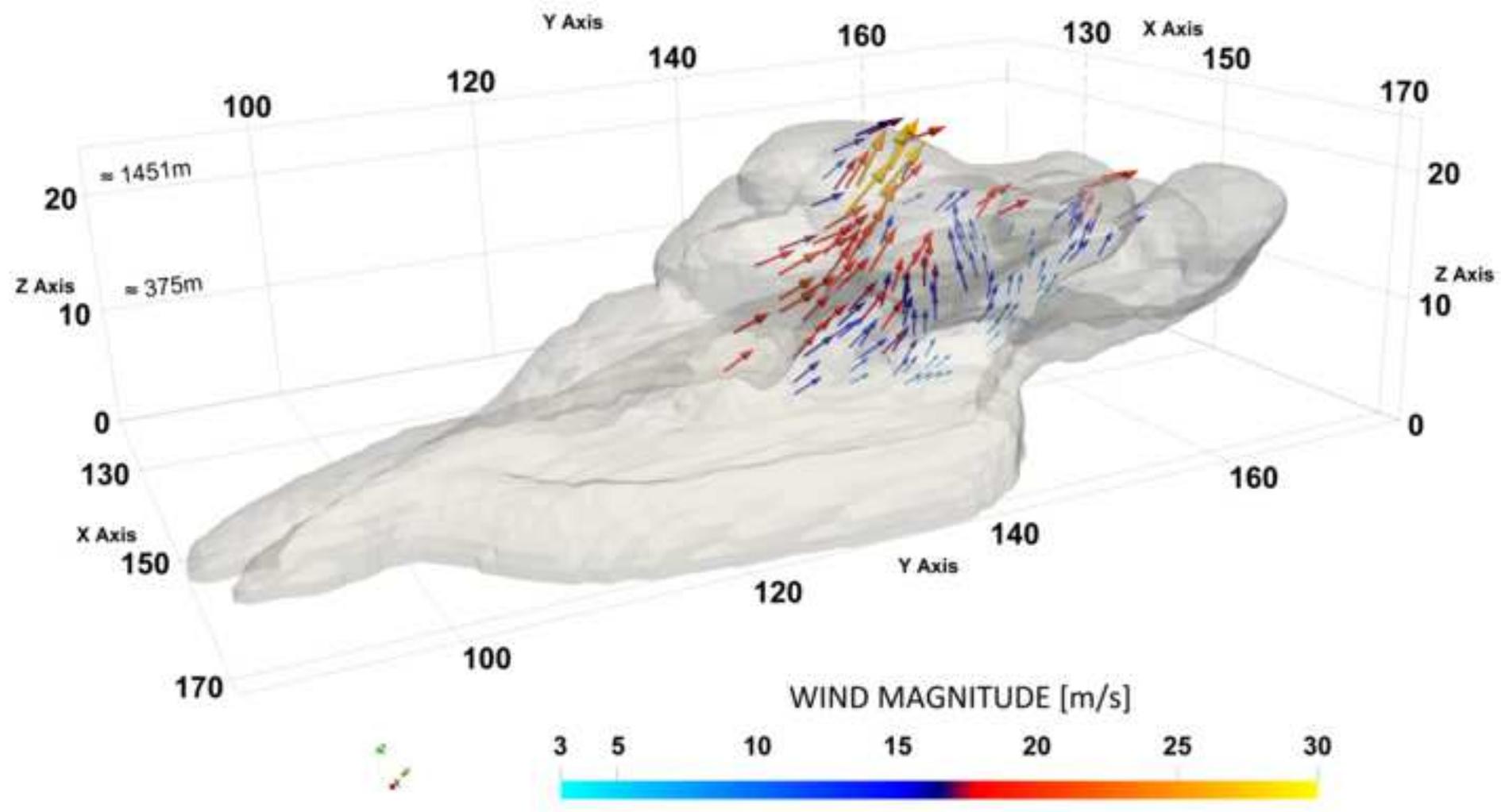

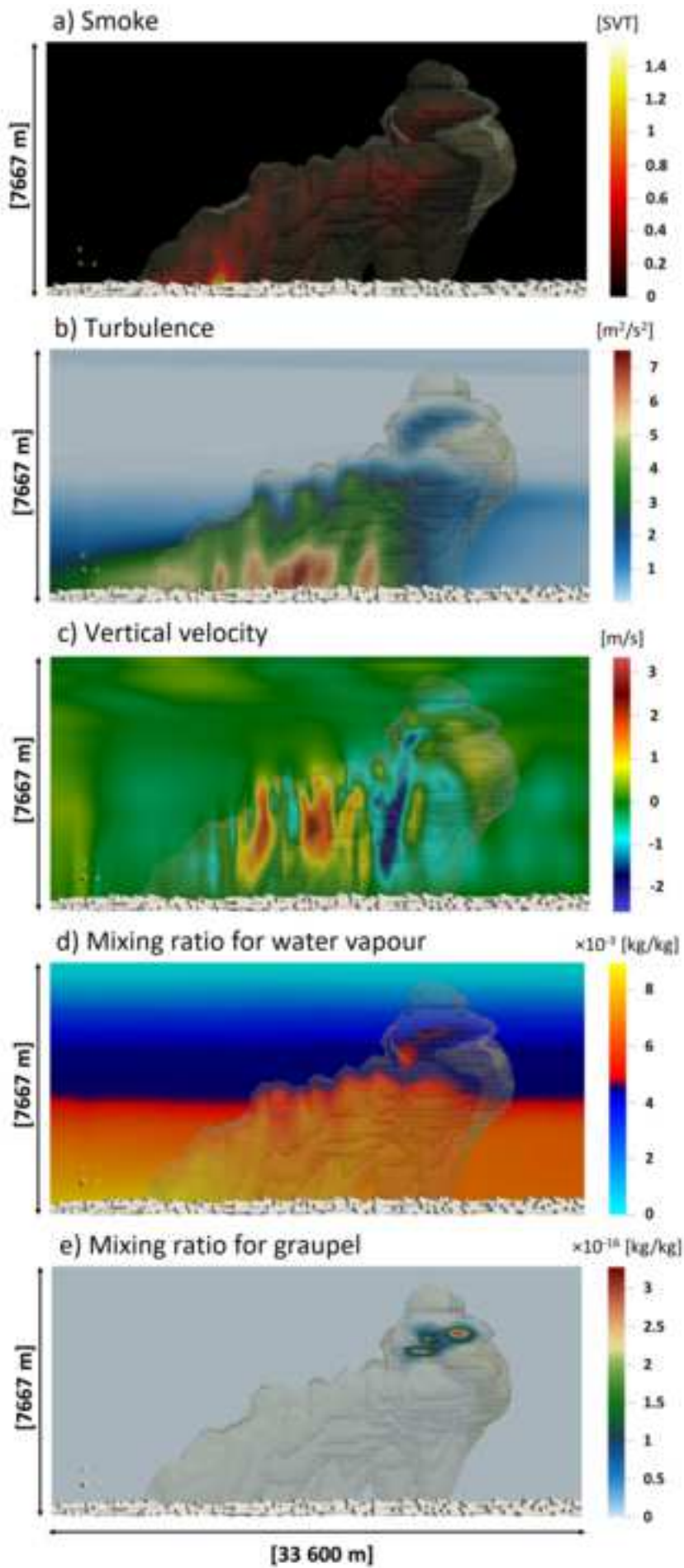

Figure_14

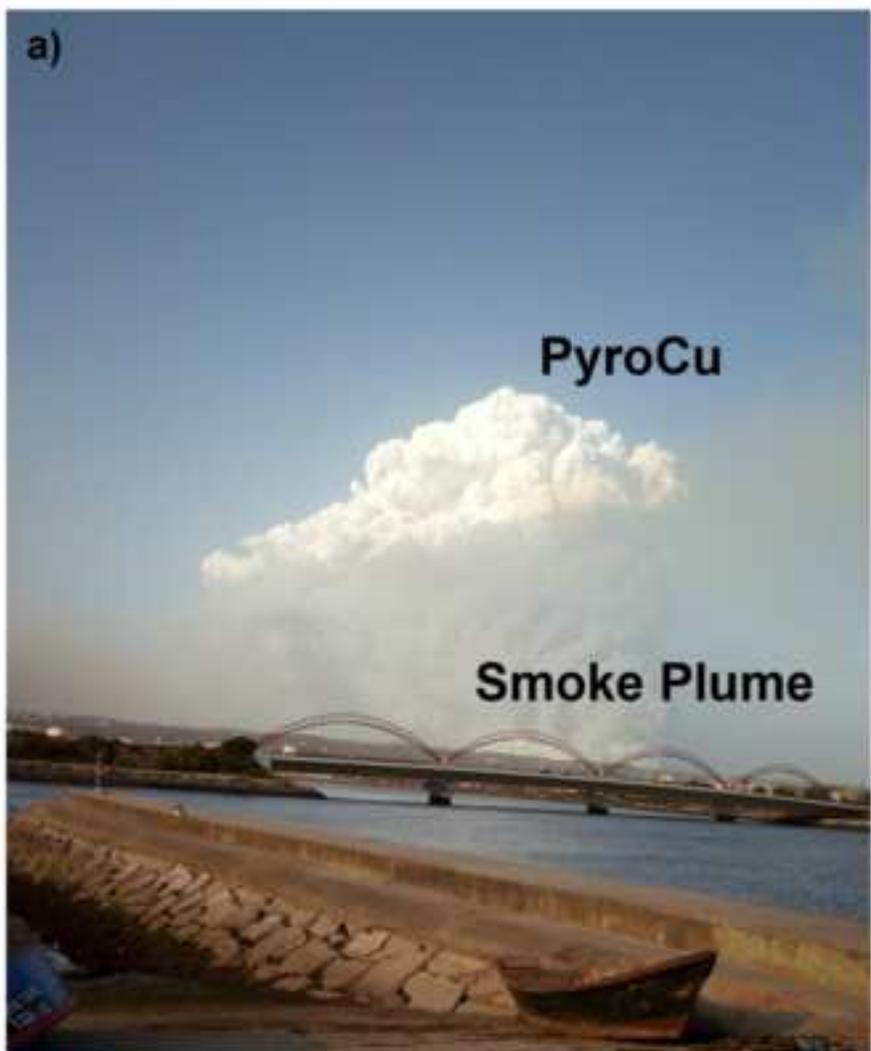
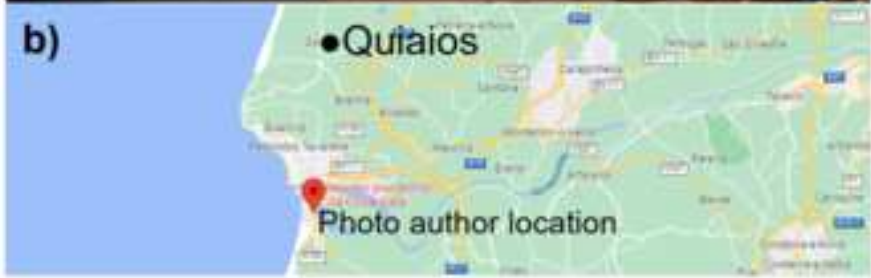
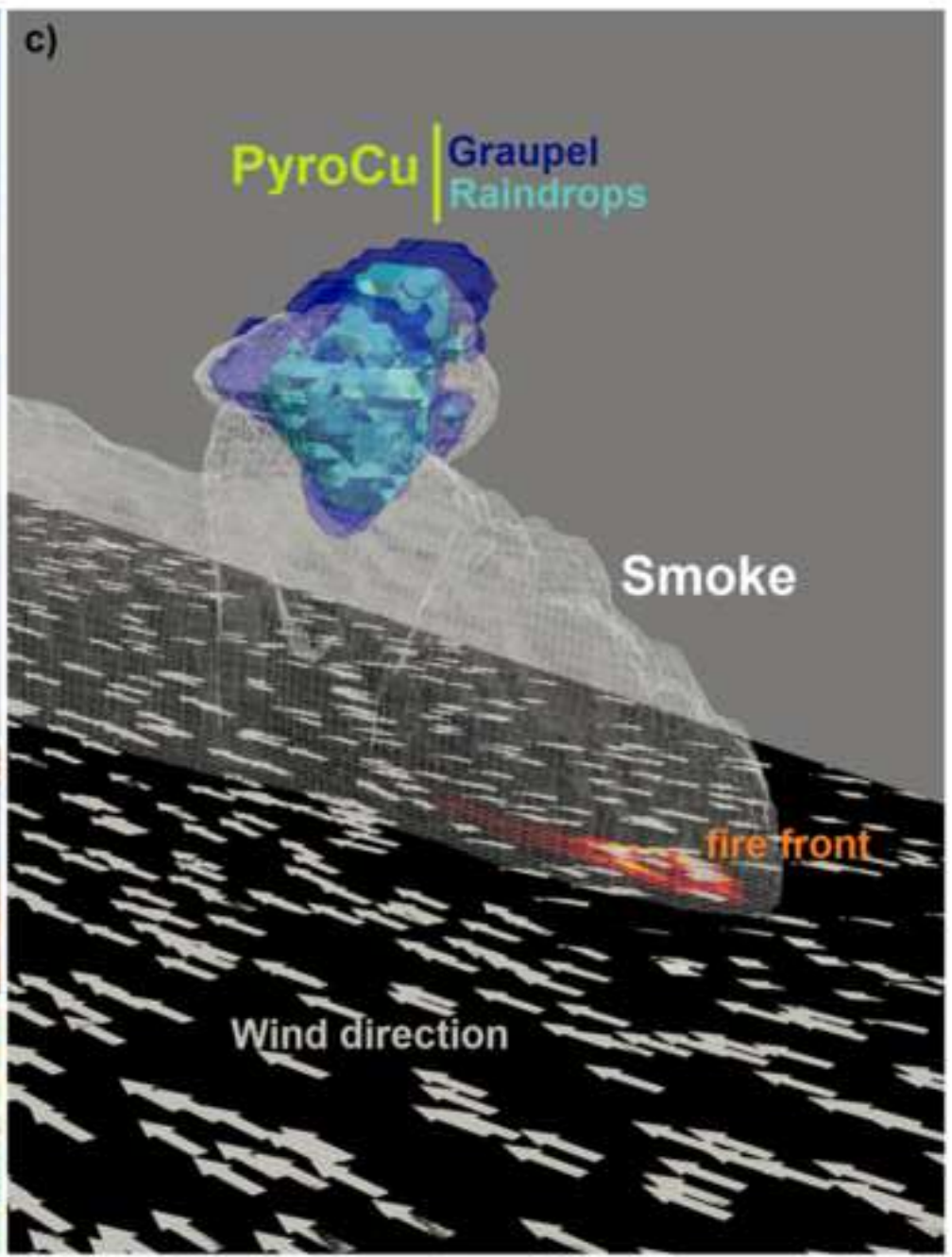

Figure_15

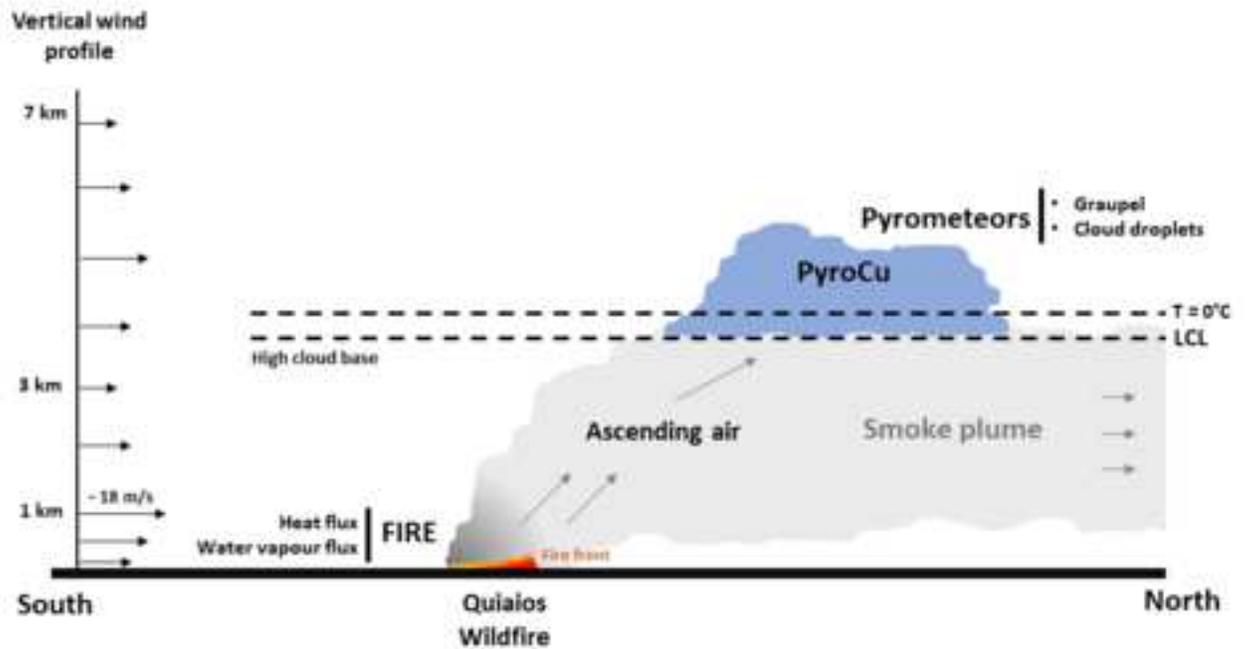

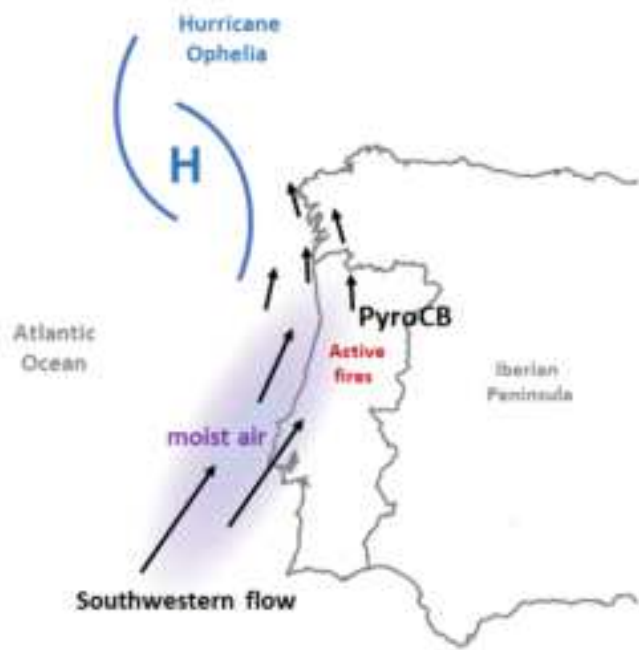

**Table 1:** Mega-fires verified over the weekend [14-16] of October 2017. Source: CTI (2018).

| District | Local | Burned area (ha) |
|---|---|---|
| **Coimbra** | Lousã – Vilarinho | 65 107.5 |
| | Arganil - Cojâ | 38 811 |
| | Figueira da Foz - Quiaios | 19 025.5 |
| **Castelo Branco** | Sertã - Figueiredo | 33 192 |
| **Viseu** | Vouzela – Campia | 22 189.8 |
| **Leiria** | Alcobaça - Pataias | 16 949.6 |
| **Guarda** | Seia - Sabugueiro | 11 924.6 |
| | Seia - Sandomil | 11 807.9 |

**Table 2:** Summary table of the parameterizations used in this study.

|  | Resolution | | | |
| --- | --- | --- | --- | --- |
| **Simulation** | **Coupled Simulation** | | | **Large - scale** |
| **Parameterization** | **2000 m** | **400 m** | **80 m** | **15 km** |
| Turbulence | 1D | 3D | 3D | 1D |
| Deep convection | ---- | ---- | ---- | KAFR |
| Shallow convection | ---- | ---- | ---- | EDKF |
| Cloud Microphysics | ICE3 | ICE3 | ICE3 | ICE3 |
| Radiation | ECMW | ECMW | ECMW | ECMW |

**Table 3:** Mean Error (ME) and Root Mean Squared Error (RMSE) for the 2m air temperature (T2M), 2m relative humidity (RH2M), and mean wind magnitude at 10 meters (WIND). The scores are calculated for the nearest point of each station in the 400 m resolution domain.

| Variable | Station | ME | RMSE |
|---|---|---|---|
| T2M | Vila Verde | 1,652166667 | 3,138159015 |
| | Dunas de Mira | -2,00841667 | 4,062462071 |
| | Becanta | -3,607 | 3,765592398 |
| | Coimbra | -3,30943333 | 3,424777179 |
| RH2M | Vila Verde | -6,37655 | 9,531444955 |
| | Dunas de Mira | 5,452383333 | 8,612037855 |
| | Becanta | 12,49298333 | 12,79472613 |
| | Coimbra | 8,669116667 | 9,248746027 |
| WIND | Vila Verde | 1,036883333 | 3,406445915 |
| | Coimbra | 0,1978 | 1,696170692 |